\documentclass[twocolumn,letter]{jpsj2}

\def\runauthor{Wataru Kobayashi}

\title{A novel heavy-fermion state in CaCu$_3$Ru$_4$O$_{12}$}

\author{Wataru {\sc Kobayashi}$^1$\thanks{E-mail address: 
kobayashi-wataru@suou.waseda.jp},
Ichiro {\sc Terasaki}$^{1,2}$,
Jun-ichi {\sc Takeya}$^3$,
Ichiro {\sc Tsukada}$^3$
and Yoichi {\sc Ando}$^3$
}

\inst{
$^1$Department of Applied Physics, Waseda University, 
Tokyo 169-8555, Japan\\
$^2$CREST, Japan Science and Technology Agency, Tokyo 108-0075, Japan\\
$^3$Central Research Institute of Electric Power Industry, 
Tokyo 201-8511, Japan
}

\date{\today}

\abst{
We have measured susceptibility, specific heat, resistivity, 
and thermopower of CaCu$_3$Ti$_{4-x}$Ru$_x$O$_{12}$ 
and CaCu$_{3-y}$Mn$_y$Ru$_4$O$_{12}$,
and have found that CaCu$_3$Ru$_4$O$_{12}$ can be regarded 
as a heavy-fermion oxide in d-electron systems. 
The Kondo temperature is near 200 K, and
the susceptibility (1.4$\times$10$^{-3}$ emu/Cu mol)
and the electron specific heat coefficient
(28 mJ/Cu molK$^2$) are moderately enhanced.
The resistivity is proportional to $T^2$ at low temperatures,
and satisfies the Kadowaki-Woods relation.
The heavy-fermion state comes from the interaction 
between the localized moment of Cu 3d 
and the conduction electron of Ru 4d. 
An insulator-metal transition occurs between $x=1.5$ and 4
in CaCu$_3$Ti$_{4-x}$Ru$_x$O$_{12}$,
which can be regarded as a transition from magnetic insulator to
heavy-fermion metal.
}

\kword{
heavy fermion, transition-metal oxide, ordered perovskite
}

\begin{document}
\maketitle

\section{\label{sec:level1}Introduction}

A large family of AC$_3$B$_4$O$_{12}$ [A=Na$^+$, Cd$^{2+}$, 
Ca$^{2+}$, Sr$^{2+}$, Y$^{3+}$, R$^{3+}$, Th$^{4+}$ or U$^{4+}$
(R; lanthanide).
B=Mn$^{3+}$, 
Fe$^{3+}$, Al$^{3+}$, Cr$^{3+}$, Ti$^{4+}$, Mn$^{4+}$, Ge$^{4+}$, 
Ru$^{4+}$, Ir$^{4+}$, Nb$^{5+}$, Ta$^{5+}$ or Sb$^{5+}$. 
C=Cu$^{2+}$ or Mn$^{3+}$]
have an ordered perovskite-type structure in which a Jahn-Teller ion 
such as Cu$^{2+}$ or Mn$^{3+}$ is located 
in the A site of the perovskite ABO$_3$. 
They have wide variety of cation compositions, and
various substitutions are possible for the A, B and C sites \cite{landolt}. 
Although they were synthesized in 1970's \cite{marezio, bochu}, 
they had not been studied so much until Subramanian {\it et al.} 
discovered an enormously large dielectric constant 
of CaCu$_3$Ti$_4$O$_{12}$ in 2000 \cite{subramanian}. 
Many researchers have investigated an origin of the large dielectric 
constant, which is still controversial at present 
\cite{ccto1, ccto4, ccto4.5, ccto5}. 
Giant magnetoresistance \cite{cmmo1} and 
phase separation \cite{cmmo2} in CaMn$_{3-x}$Cu$_x$Mn$_4$O$_{12}$ 
are a second example. 
We think that these interesting physical properties 
are related to the peculiar crystal structure, 
and have studied the transport properties of 
the AC$_3$B$_4$O$_{12}$ compounds\cite{ccto4, cmmo3, cmmo4}.

A heavy-fermion system is a strongly correlated electron system where 
localized f-electrons interact with conduction electrons through the Kondo effect. 
In this system, a high and narrow peak appears in the quasiparticle 
density of states $D$ near the Fermi energy $E_F$. 
The large $D(E_F)$ is reflected in a large susceptibility 
and electron specific-heat coefficient. 
LiV$_2$O$_4$  has been regarded as a typical example
of the heavy fermion system in d-electron systems \cite{LiVO}. 
In LiV$_2$O$_4$, however, V 3d electrons act as 
conduction electrons and local moments simultaneously. 
Thus we cannot distinguish the local moment site 
from the conduction electron site. 
In this meaning, LiV$_2$O$_4$ is 
not equivalent to the f-electron heavy-fermion system.

CaCu$_3$Ru$_4$O$_{12}$ is one of the AC$_3$B$_4$O$_{12}$-type
oxides, and exhibits a good metallic conduction down to 4 K.
It shows higher conductivity than CaRuO$_3$ with less tilted RuO$_6$
octahedra, which would be difficult to understand 
if only the Ru-O network would be conductive.
Subramanian and Sleight \cite{ccro1} proposed 
``valence degeneracy'' for the electronic states of
CaCu$_3$Ru$_4$O$_{12}$, where Cu$^{2+}$ would contribute to 
the electric conduction.
Their explanation was, however, phenomenological,
and the microscopic mechanism was not clearly understood. 
In this paper, we will propose that CaCu$_3$Ru$_4$O$_{12}$ is 
a novel d-electron heavy-fermion system, 
which is truly equivalent to the f-electron heavy-fermion system
in the sense that the local moment (Cu 3d)
and the conduction electron (Ru 4d)
are clearly distinguished. 
This is evidenced by the substitution of Mn for Cu
in CaCu$_{3-y}$Mn$_y$Ru$_4$O$_{12}$,
where the anomalous impurity effects are observed.
We have further found that an insulator-metal transition occurs in 
CaCu$_3$Ti$_{4-x}$Ru$_x$O$_{12}$, which can be regarded as a 
transition from magnetic insulator to heavy-fermion metal.

\section{Experimentalal}

Polycrystalline samples of CaCu$_3$Ti$_{4-x}$Ru$_x$O$_{12}$ 
($x=$ 0, 0.5, 1 and 1.5), 
and CaCu$_{3-y}$Mn$_y$Ru$_4$O$_{12}$ 
($y=$ 0, 0.1, 0.2 and 0.3) were prepared by a solid-state reaction
(Note that the sample of $y=0$ is identical to that of $x=4$). 
Stoichiometric amounts of CaCO$_3$, CuO, RuO$_2$, 
TiO$_2$, and Mn$_3$O$_4$ were mixed, 
and  CuO flux was added to 
CaCu$_{3-y}$Mn$_y$Ru$_4$O$_{12}$ \cite{gousei}. 
The mixture was calcined in air
at 1000-1050 $^{\circ}$C for 20 h for CaCu$_3$Ti$_{4-x}$Ru$_x$O$_{12}$, 
1050 $^{\circ}$C for 48 h for 
CaCu$_{3-y}$Mn$_y$Ru$_4$O$_{12}$. 
The CuO flux in the latter was removed by 1N HCl.
The product was finely ground, pressed into a pellet, 
and sintered in air at 1000-1050 $^{\circ}$C for 20 h for 
CaCu$_3$Ti$_{4-x}$Ru$_x$O$_{12}$, 
1050 $^{\circ}$C for 24 h for CaCu$_{3-y}$Mn$_y$Ru$_4$O$_{12}$.

The X-ray diffraction was measured using 
a standard diffractometer with Cu K$_\alpha$ radiation as an X-ray source 
in the $\theta$-$2\theta$ scan mode. 
The susceptibility was measured from 5 to 300 K 
by a commercial Superconducting Quantum Interference Device 
(SQUID, Quantum Design MPMS) in 1 T. 
The specific heat was measured using a standard relaxation method 
from 2 to 60 K. 
The resistivity was measured by the four-probe method
from 4.2 to 300 K in a liquid He cryostat. 
For CaCu$_3$Ru$_4$O$_{12}$, we extended the measurement down to
0.3 K using a $^3$He cryostat.
but did not find any indication of superconductivity.
The thermopower was measured using a steady-state technique
from 4.2 to 300 K in a liquid He cryostat. 
A temperature gradient of 0.5 K/mm was generated 
by a small resistive heater pasted on one edge of a sample, 
and was monitored by a differential thermocouple made 
of copper-constantan. 
The thermopower of the voltage leads was carefully subtracted.

\begin{figure}[b]
 \begin{center}
  \includegraphics[width=6cm,clip]{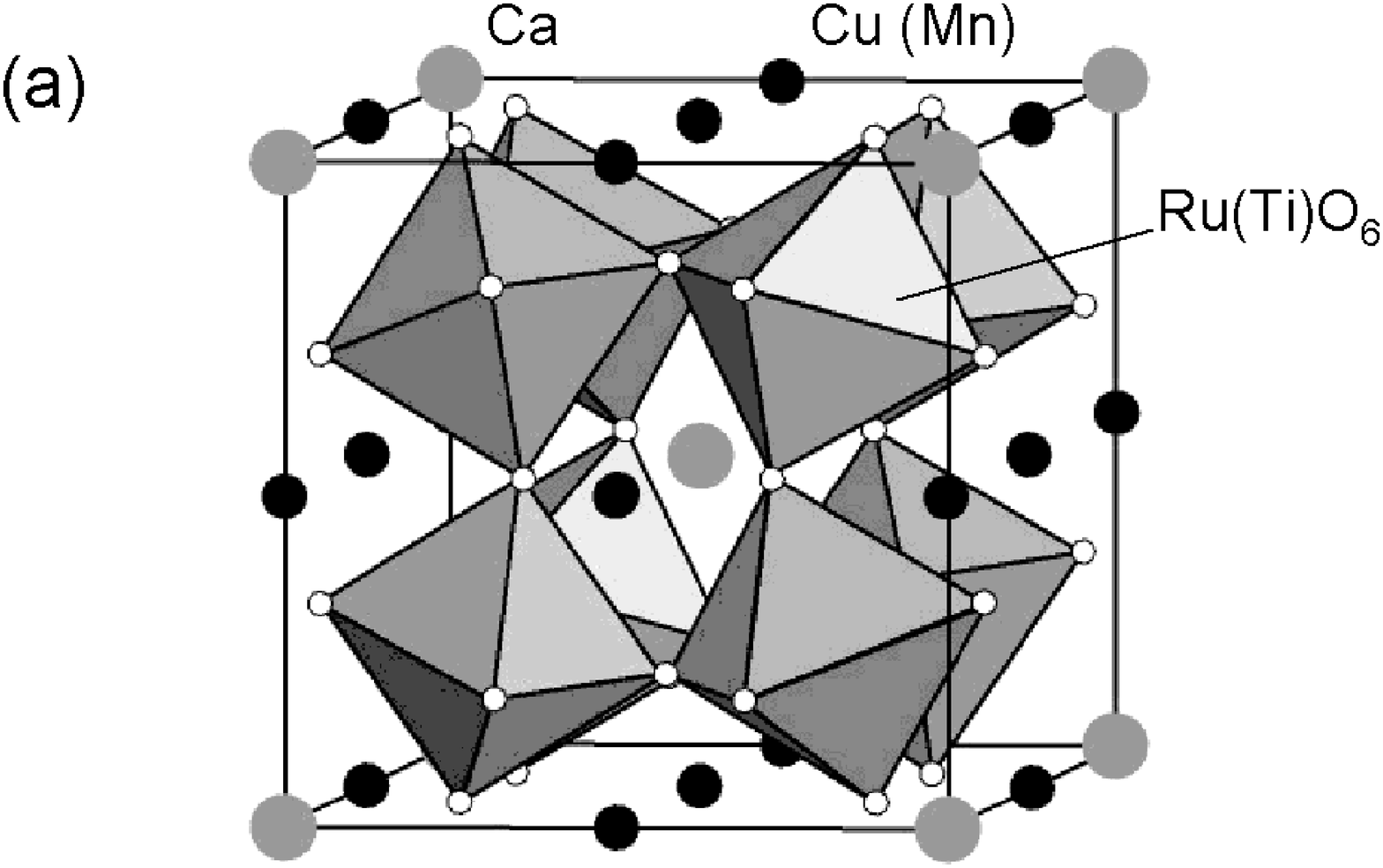}
 \end{center} 
 \begin{center}
  \includegraphics[width=8cm,clip]{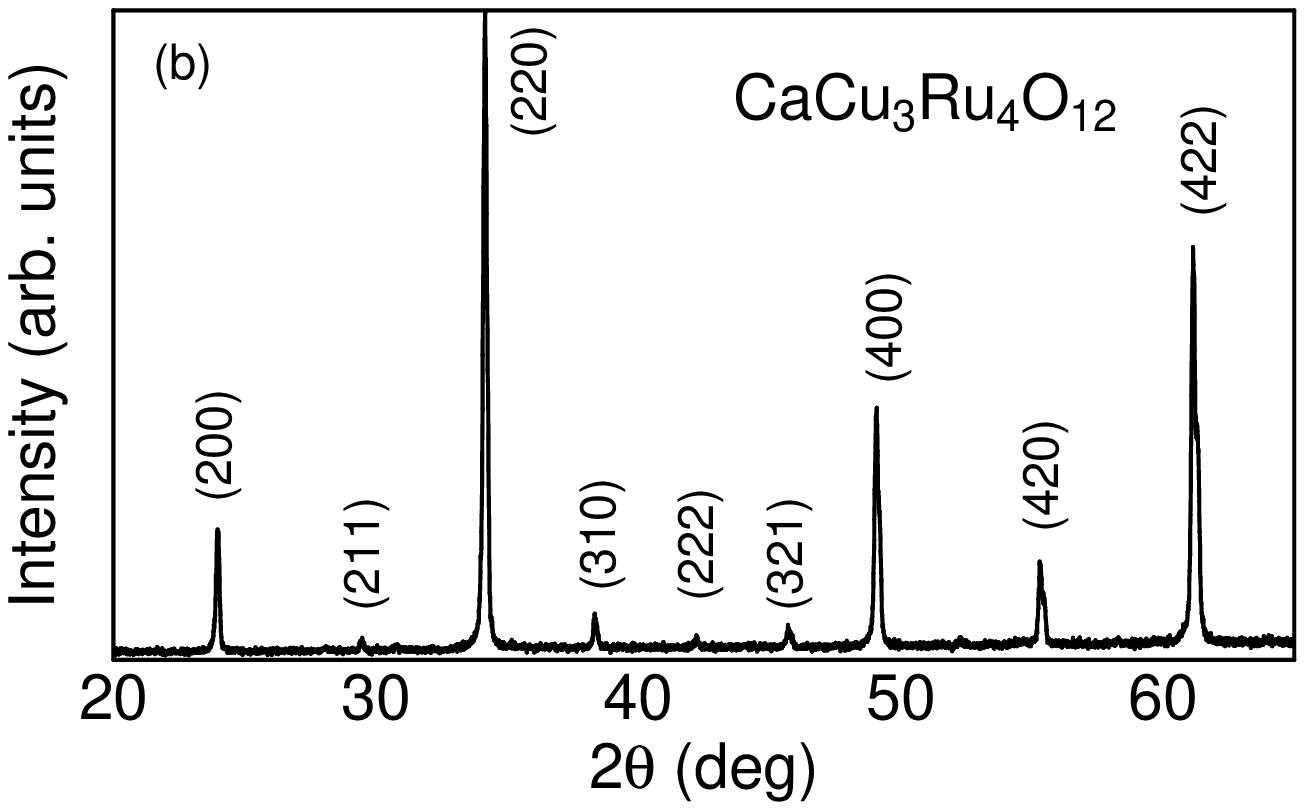}
 \end{center} 
 \caption{
 (a) Crystal structure  and 
 (b) X-ray diffraction pattern of CaCu$_3$Ru$_4$O$_{12}$.
 }
\end{figure}

\section{Results and Discussion}

Figure 1(a) shows the crystal structure of CaCu$_3$Ru$_4$O$_{12}$,
which is the AC$_3$B$_4$O$_{12}$ type structure with Cu located at the C site. 
Since Cu$^{2+}$ is a small ion, 
the RuO$_{6}$ octahedra are highly canted to make 
the lattice parameter smaller than that of 
CaRuO$_3$ or SrRuO$_3$.
Another interesting feature is that the Cu-O distance (1.94 \AA) 
is as short as the  Ru-O distance (1.98 \AA) \cite{gousei}.
Cu is surrounded with four oxygens that are shared with the RuO$_6$ octahedra, 
and thus we expect a strong hybridization between Cu 3d and O 2p
and between Ru 4d and O 2p, through which Ru 4d can interact with Cu 3d.
As is shown in Fig. 1 (b), the X-ray diffraction pattern of 
CaCu$_3$Ru$_4$O$_{12}$ is fully indexed 
as the AC$_3$B$_4$O$_{12}$ structure \cite{laveau}. 
Other samples are also fully indexed 
on the structure reported in the literature \cite{laveau,bochu} 
except for 5\% RuO$_2$ impurity 
in CaCu$_3$Ti$_{2.5}$Ru$_{1.5}$O$_{12}$.

\begin{figure}[t]
 \begin{center}
  \includegraphics[width=8cm,clip]{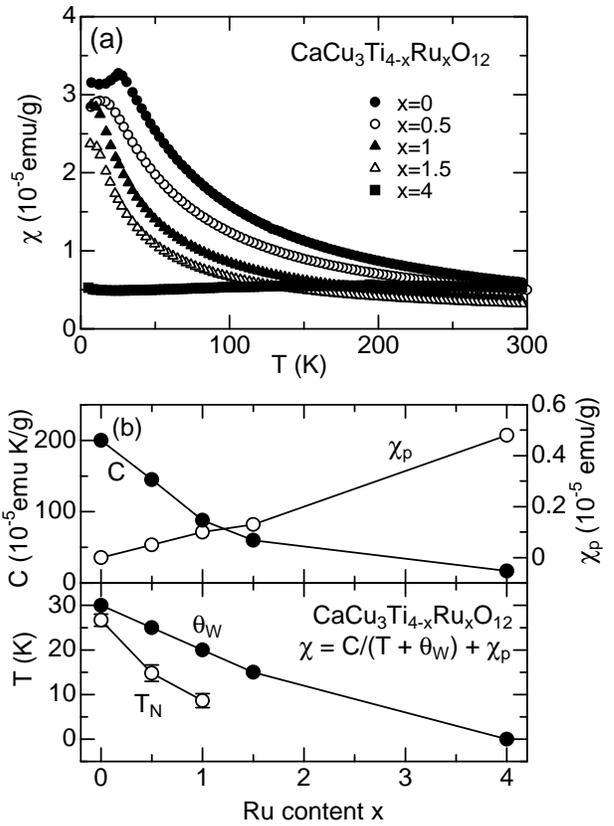}
 \end{center} 
 \caption{
 (a) The magnetic susceptibility of 
 CaCu$_3$Ti$_{4-x}$Ru$_x$O$_{12}$
 ($x$=0, 0.5, 1, 1.5 and 4) 
 and (b) the $x$ dependences of the Curie constant ($C$), 
 the Pauli paramagnetic susceptibility ($\chi_p$),  
 the Weiss temperature ($\theta_W$) and the 
 Neel temperature ($T_N$).
 }
\end{figure}

Figure 2 (a) shows the susceptibility of CaCu$_3$Ti$_{4-x}$Ru$_x$O$_{12}$.
We have found that the high-temperature data are well fitted 
with the expression $\chi (T)=\chi_p+\chi_{loc}(T)$, where 
$\chi_p$ is the Pauli paramagnetic susceptibility, and $\chi_{loc}$ is the 
Curie-Weiss-type susceptibility $\chi_{loc}(T)=C/(T+\theta_W$)
($C$ is the Curie constant, and $\theta_W$ is the Weiss temperature). 
Figure 2 (b) shows the parameters of $C$, $\chi_p$ and $\theta_W$. 
$C$ for $x$=0 is 2$\times$10$^{-3}$ emu K/g, 
which corresponds to 1.3 $\mu_B$/Cu. 
This indicates that Cu$^{2+}$ for $x=0$ acts as a local spin moment of $S=1/2$.
With increasing $x$, $C$ rapidly decreases down to 6$\times$10$^{-4}$ 
emu K/g (0.4 $\mu_B$/Cu) for $x=1.5$, 
and instead $\chi_p$ linearly increases. 
An important feature is that the susceptibility for all the samples 
seems to merge into a single curve at high temperatures,
which suggests that the sum of the contributions of $\chi_{loc}$ and 
$\chi_p$ is conserved.
This strongly suggests that the localized moments on Cu$^{2+}$
becomes itinerant with increasing $x$.
$\chi$ of $x=4$ was fitted only above 200 K,
where $C$ is 1.7 $\times$10$^{-4}$ emu K/g (0.1 $\mu_B$/Cu).
It weakly depends on $T$ below 200 K, 
which will be discussed in the next paragraph.
$\theta_W$ and the Neel temperature $T_N$ for $x$=0 are 30 and 27  K, 
respectively, which is consistent with the previous data \cite{ccto4.5}.
$T_N$ decreases with $x$ and disappears near $x=2$,
which is much faster than the temperatures expected from the dilution effect.

\begin{figure}[t]
 \begin{center}
  \includegraphics[width=8cm,clip]{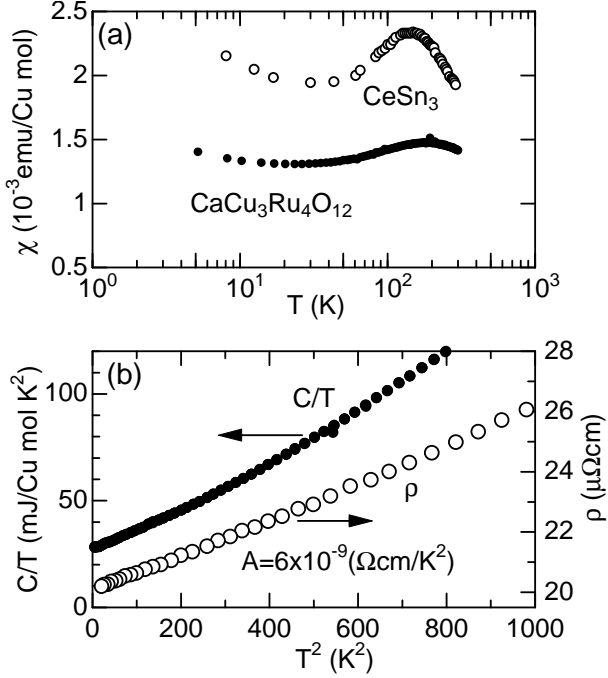}
 \end{center} 
 \caption{
 (a) The susceptibility $\chi$ and  
 (b) the specific heat ($C$) and the resistivity $\rho$ of CaCu$_3$Ru$_4$O$_{12}$.
 Note that $\chi$ of CeSn$_3$ is plotted in a unit of emu/Ce mol \cite{CeSn}.
 }
\end{figure}

As mentioned earlier, the microscopic mechanism of 
the delocalization of the Cu$^{2+}$ moment seen in Fig. 2
was not yet understood.
We propose that the localized Cu$^{2+}$ moment starts to be itinerant
through the Kondo coupling with Ru 4d,
and finally forms a heavy-fermion state.
Here we will show convincing evidence for the 
heavy-fermion state of CaCu$_3$Ru$_4$O$_{12}$. 
First, CaCu$_3$Ru$_4$O$_{12}$ shows a typical susceptibility
of heavy-fermion metals.
As is shown in Fig. 3 (a),
$\chi$ of CaCu$_3$Ru$_4$O$_{12}$ 
quite resembles that of CeSn$_3$ \cite{CeSn}, a typical heavy-fermion metal.
A value of 1.4$\times10^{-3}$emu/Cu mol is two orders of magnitude
larger than the value of free electrons, and 
a peak at 200 K is an indication of the Kondo resonance. 
These results strongly indicate that the Ru 4d electron 
interacts with the localized Cu moment to make a heavy-fermion state 
below 200 K. 
The second piece of the evidence is a large electron specific heat.
Figure 3 (b) shows the specific heat of CaCu$_3$Ru$_4$O$_{12}$. 
The electron specific heat coefficient ($\gamma$) is 
observed to be 28 mJ/Cu molK$^2$ 
that is about 20-30 times larger than the value of free electrons. 
The Willson ratio $R_W\equiv\pi^2k_B^2\chi_p/3\mu_B^2\gamma$ 
is 3.8, which is in an excellent agreement with other heavy-fermion 
compounds.
Thirdly, the resistivity exhibits typical features of heavy fermion metals.
As shown in Fig. 3(b), the resistivity ($\rho$) of CaCu$_3$Ru$_4$O$_{12}$ 
is plotted as a function of $T^2$, which clearly shows the relation
of $\rho=AT^2+\rho_0$ as is expected in the Fermi liquid. 
The coefficient $A$ is evaluated to be 6$\times$10$^{-9}$ $\Omega$cm/K$^2$, 
which satisfies the Kadowaki-Woods relation:
$A/\gamma^2$ is nearly the same value as that for heavy-fermion metals.

\begin{figure}[t]
 \begin{center}
  \includegraphics[width=7cm,clip]{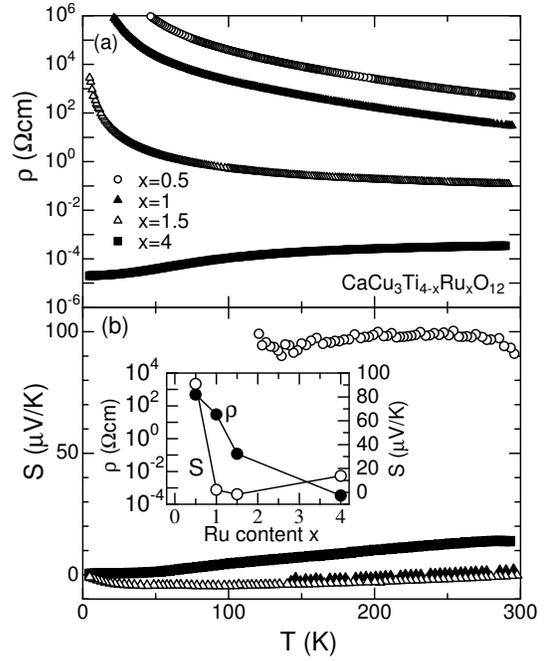}
 \end{center} 
 \caption{
 (a) The resistivity $\rho$, (b) the thermopower $S$ of 
 CaCu$_3$Ti$_{4-x}$Ru$_x$O$_{12}$
 ($x$=0, 0.5, 1, 1.5 and 4).
 The inset: the $x$ dependence of $\rho$ and $S$ at 300 K.
 }
\end{figure}

We would like to mention some novelties of the heavy-fermion state
in CaCu$_3$Ru$_4$O$_{12}$.
First of all, this is the first compound in which Cu$^{2+}$ ($3d^9$) 
behaves like Ce$^{3+}$ ($4f^1$) to make conduction electrons heavy.
This comes from the peculiar structure that RuO$_2$ octahedra
and the Cu-O placket are connected in a edge-shared manner.
Second, this can be regarded as a conductive perovskite
in which the A site is responsible for electric conduction.
This is so rare that 
BiNiO$_3$ would be the only one example ever reported \cite{BiNi},
which is insulating at low temperatures. 
Thirdly, the conduction electrons are Ru 4d, 
which can be also magnetic.
As is well known, the perovskite oxide SrRuO$_3$ 
is a ferromagnetic metal.
It is thus surprising that the ground state of CaCu$_3$Ru$_4$O$_{12}$
is ``nonmagnetic'', though almost all cations are magnetic.

Figure 4(a) shows $\rho$  of 
CaCu$_3$Ti$_{4-x}$Ru$_x$O$_{12}$,
which changes from 10$^3$ ($x=0$) to 
10$^{-4}$ $\Omega$cm ($x=4$) at 300 K.  
The temperature dependence also changes from 
insulating to metallic, and 
an insulator-metal transition occurs between $x=1.5$ and $x=4$.
Figure 4(b) shows the thermopower ($S$) of 
CaCu$_3$Ti$_{4-x}$Ru$_x$O$_{12}$. 
Most unexpectedly, $S$ suddenly decreases
from 100 $\mu$V/K for $x=0.5$ to a few $\mu$V/K for $x=1.0$,
whereas $\rho$ for $x$=1.0 is still high.
Since the small thermopower is a hallmark of a metal, 
this indicates that the delocalization of the 3d holes on Cu$^{2+}$,
{\it i.e.}, the Kondo coupling with the Ru 4d electron,
already occurs at $x=1$.
We infer from the Curie constant in Fig. 2(b)
that a half of the 3d holes are coupled with Ru 4d.
Although the Kondo coupling between Ru 4d and Cu 3d exists,
the conduction path is seriously segmentized at $x=1$.
This is the reason of the high resistivity for $x=1$.
In contrast, the thermopower is a direct probe
for the entropy per {\it conductive carriers},
it is less affected by the insulating region in the sample.
In fact, $E_F=$1.4 eV estimated for 50\% of Cu$^{2+}$
gives the diffusive term of the thermopower 
$k_B^2T/eE_F=$1 $\mu$V/K at 300 K,
which is consistent with the observed data.
We further note that the thermopower of CaCu$_3$Ru$_4$O$_{12}$ is similar to
that of CeSn$_3$ \cite{mahan},
which is consistent with the data in Fig. 2.

Figure 5(a) shows $\rho$ of 
CaCu$_{3-y}$Mn$_y$Ru$_4$O$_{12}$. 
$\rho$ systematically increases with $y$, which further supports
that the Cu site contributes to the conduction. 
$\rho$ for the Mn-doped samples is understood in terms of
the sum of $\rho$ for $y=0$ and the impurity-scattering term
induced by Mn. 
The impurity scattering is usually independent of temperature,
and the resistivity of an impurity-doped sample is shifted upward 
in parallel to the resistivity of a pure sample. 
However, this is not the case.
The residual resistivity at 4 K is roughly proportional to the Mn content,
as is seen in the impurity scattering of a conventional metal.
On the other hand, the resistivity is nearly independent of $y$
at room temperature.
This can be also understood from the heavy-fermion scenario.
At room temperature (above the Kondo temperature), 
Mn$^{3+}$ and Cu$^{2+}$ seem similar magnetic scatterers 
to the conduction electrons.
In contrast, the Cu 3d moments coherently move with the conduction
electrons below the Kondo temperature, where
only the Mn$^{3+}$ remains as a scatterer.

\begin{figure}[t]
 \begin{center}
  \includegraphics[width=7cm,clip]{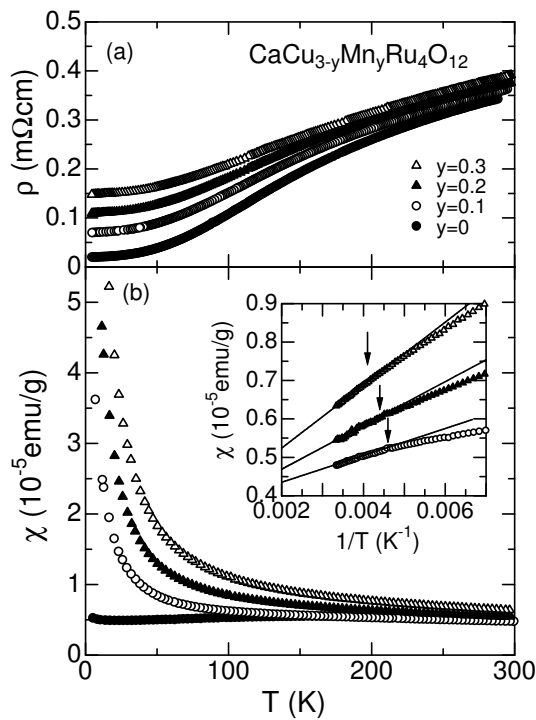}
 \end{center} 
 \caption{
 (a) The resistivity $\rho$ and (b) the susceptibility $\chi$ of 
 CaCu$_{3-y}$Mn$_y$Ru$_4$O$_{12}$ 
 ($y$=0, 0.1, 0.2 and 0.3).
 }
\end{figure}

Figure 5(b) shows $\chi$ of
CaCu$_{3-y}$Mn$_y$Ru$_4$O$_{12}$, 
which obeys the Curie law for $y>0$. 
The Curie constant for $y>0$ corresponds to 2 $\mu_B$/Mn, 
which is a half of the value expected from 
Mn$^{3+}$ ($S=2$). 
We do not understand the reason of the unusual small moment of Mn$^{3+}$.
One possibility is that Mn$^{3+}$ is antiferromagnetically
coupled with the neighboring Cu$^{2+}$. 
As shown in the inset, $\chi$ exhibits a kink near 200 K
which means that the Kondo coupling remains against the Mn doping.
This is consistent with the metallic nature of the Mn-doped samples.

\section{Summary}
The susceptibility, electron specific heat,
resistivity and thermopower of CaCu$_3$Ru$_4$O$_{12}$
are consistently and quantitatively understood as
those of a heavy-fermion metal with the Kondo temperature of 200 K.
This comes from the interaction between Cu 3d and Ru 4d electrons.
The former corresponds to the localized $f$ electron,
and the latter corresponds to the conduction electron.
In this sense, CaCu$_3$Ru$_4$O$_{12}$ is a complete analogue to the 
f-electron heavy fermion metal.
We have further found that a insulator-metal transition in 
CaCu$_3$Ti$_{4-x}$Ru$_x$O$_{12}$, which can be regarded as a
transition from magnetic insulator to heavy fermion state,
through which the hole on Cu$^{2+}$ starts to be itinerant 
by the help of the Kondo coupling with Ru 4d.

Before completion of the present manuscript, we found the preprint of 
Ramirez et al. on the preprint archives (cond-mat/0403742).
They discussed the itinerancy of the Cu$^{2+}$ moment in
LaCu$_3$Ti$_{4-x}$Ru$_x$O$_{12}$.
However, they did not show susceptibility data
of CaCu$_3$Ru$_4$O$_{12}$, 
and did not mention the heavy-fermion state either.

\begin{acknowledgments}
The authors would like to thank Shintaro Ishiwata for fruitful
discussion on BiNiO$_3$.
This work was partially supported by Grant-in-Aid for JSPS Fellows.
\end{acknowledgments}

\end{document}